\documentclass[12pt]{article}
\begin{document}

\begin{title}
{All homogeneous pure radiation spacetimes satisfy the Einstein-Maxwell equations}
\end{title}

\author{
C. G. Torre\\ {\sl Department of Physics}\\
{\it Utah State University}\\ 
{\it Logan, UT, USA, 84322-4415}}
\maketitle

\begin{abstract}
It is shown that all homogeneous pure radiation solutions to the Einstein equations admit  electromagnetic sources. This corrects an error in the literature.
\end{abstract}
\bigskip\bigskip

A spacetime $(M, g)$ is said to be a homogeneous pure radiation solution of the Einstein equations if it admits a transitive group of isometries and if there exists a function $\Phi$ and 1-form $k$ such that the Einstein tensor $G$ is of the form
\begin{equation}
G_{ab} = \Phi^2 k_a k_b,\quad{\rm where}\quad g^{ab}k_a k_b = 0.
\end{equation}
A classification of homogeneous pure radiation solutions is given in Ref.~\cite{SKMHH} based upon results of  Refs.~\cite{Wils} and \cite{Steele}. Such solutions are all pp waves, and they are either plane waves or they are diffeomorphic to
\begin{equation}
g = dx \otimes dx + dy \otimes dy + du \odot dv - 2 e^{2\rho x} du \otimes du,\quad \rho= const.
\end{equation}
The plane waves are known to arise from electromagnetic sources.  
It is asserted in Refs.~\cite{SKMHH, Steele}, based upon results of Ref.~\cite{SG},  that the metric (2) does not arise from an electromagnetic source. However, the metric (2) {\it does} arise from an electromagnetic source. In fact, infinitely many electromagnetic fields can serve as the source for the spacetime defined by (2).

For any function $f(u)$ the null two form 
\begin{equation}
F = 2\rho e^{\rho x} \Bigg[\cos(\rho y + f(u))\, du \wedge dx - \sin(\rho y + f(u))\,du \wedge dy\Bigg] ,
\end{equation}
its Hodge dual $\tilde F$, and the metric (2) satisfy the source-free Maxwell equations $dF = 0 = d \tilde F$ and the Einstein equations
\begin{equation}
G_{ab} = \frac{1}{2}\left(F_{ac}F_b{}^c + \tilde F_{ac} \tilde F_b{}^c\right) = 4\rho^2 e^{2\rho x} \nabla_a u  \nabla_b u.
\end{equation}
The electromagnetic field given in (3) is ``non-inheriting'', that is, it is not invariant under the  isometry group of $(M, g)$.  For example, $F$ is not invariant under translations in $y$  for any choice of $f(u)$.

Thus, granted Theorem 12.6 of \cite{SKMHH}, all homogeneous pure radiation solutions of the Einstein equations arise from  electromagnetic sources. 

\bigskip\noindent
{\bf Acknowledgments:} I thank Ian Anderson for discussions. These results were obtained using the {\sl DifferentialGeometry} package in {\sl Maple}. 

{}

\end{document}